\documentclass[prl,twocolumn,amsmath,amssymb,10pt,aps,longbibliography,superscriptaddress,citeautoscript,bibnotes,footinbib]{revtex4-2}

\usepackage[normalem]{ulem} 

\usepackage{feynmp}
\usepackage{amsmath}
\usepackage{graphicx}
\usepackage{multirow}
\usepackage{latexsym}
\usepackage{textcomp}
\usepackage{verbatim}
\usepackage{color}
\usepackage{bm}
\usepackage{subfigure}
\usepackage{everysel}
\usepackage{keyval}
\usepackage{ragged2e}
\usepackage{dsfont}
\usepackage{amssymb}
\usepackage{enumitem}
\usepackage{pstricks}
\usepackage{mathtools}
\usepackage{braket}
\usepackage{slashed}
\usepackage[colorlinks=true]{hyperref}

\usepackage{graphicx}
\usepackage{xcolor}
\usepackage{amsmath}
\usepackage{amsfonts}
\usepackage{bbm}

\newcommand{\osum}{{%
    \setbox0\hbox{\circ}%
    \rlap{\hbox to \wd0{\hss\sum\hss}}\box0
}}

\begin{document}

\title{Non-Fermi Liquids in Conducting 2D Networks}

\author{Jongjun M. Lee}
\thanks{Electronic Address: michaelj.lee@postech.ac.kr}
\affiliation{Department of Physics, Pohang University of Science and Technology (POSTECH), Pohang 37673, Republic of Korea}

\author{Masaki Oshikawa}
\thanks{Electronic Address: oshikawa@issp.u-tokyo.ac.jp}
\affiliation{Institute for Solid State Physics, The University of Tokyo, Kashiwa 277-8581, Japan}
\affiliation{Kavli Institute for the Physics and Mathematics of the Universe, Kashiwa 277-8583, Japan }
\affiliation{Trans-scale Quantum Science Institute, University of Tokyo, Bunkyo-ku, Tokyo 113-0033, Japan}

\author{Gil Young Cho}
\thanks{Electronic Address: gilyoungcho@postech.ac.kr}
\affiliation{Department of Physics, Pohang University of Science and Technology (POSTECH), Pohang 37673, Republic of Korea}
\affiliation{Center for Artificial Low Dimensional Electronic Systems,  Institute for Basic Science (IBS),  Pohang 37673, Korea}

\date{\today}

\begin{abstract}
We explore the physics of novel fermion liquids emerging from conducting networks, where 1D metallic wires form a periodic 2D superstructure. Such structure naturally appears in marginally-twisted bilayer graphenes, moire transition metal dichalcogenides, and also in some charge-density wave materials. For these network systems, we theoretically show that a remarkably wide variety of new non-Fermi liquids emerge and that these non-Fermi liquids can be \textit{classified} by the characteristics of the junctions in networks. Using this, we calculate the electric conductivity of the non-Fermi liquids as a function of temperature, which show markedly different scaling behaviors than a regular 2D Fermi liquid.
\end{abstract}

\maketitle

{\color{green} \textbf{\textit{1. Introduction}:}} The ubiquity of Fermi liquids~\cite{Nozieres-Pines} makes it particularly interesting to find and elucidate systems in which the Fermi liquid theory breaks down, namely \textit{non-Fermi liquid} (NFL) behavior.
Experimentally, one of the signatures of NFL is the resistivity exponent $\alpha$ defined the resistivity
\begin{align}
  \rho_{xx}(T) = \rho_0 + c~ T^{\alpha} ,
  \label{eq.resistivity_power}
\end{align}
as a function of the temperature $T$.
The Fermi liquid is characterized by $\alpha = 2$. 
On the other hand, various values of the exponent $\alpha \neq 2$ been observed experimentally in strongly correlated electron systems, indicating NFL.
However, a controlled theoretical description of NFLs in $d \geq 2$ dimensions remains a challenging problem~\cite{lee2018recent, schofield1999non, varma2002singular}.
In this Letter, we propose a simple theoretical model of NFL in $d \geq 2$ dimensions at finite temperatures, including ``strange insulator'' with a negative exponent $\alpha$\cite{donos2013interaction, donos2014novel, andrade2019coherent}, in terms of a network made of 1D conducting segments.

While our theory is specific to network superstructures, such systems appear in a surprisingly wide variety of materials.
To name only a few, marginally twisted bilayer graphene~\cite{rickhaus2018transport, yoo2019atomic, xu2019giant}, moire transition metal dichalcogenides~\cite{ACS, carr2018relaxation,weston2020atomic},
helium atoms absorbed on graphene~\cite{Morishita2019}, and certain charge-density wave materials~\cite{Park-Prep, NC-CDW-Xray,li2016controlling} show such superstructures. Possibility of engineering a network in ultra cold gas experiments is discussed in \cite{suppl_ref}.
A series of intriguing many-body phenomena have been observed in these systems,
including superconductivity~\cite{sipos2008mott, yu2015gate, PhysRevB.94.045131,
li2016controlling,chen2019discommensuration,wang2019magic,kusmartseva2009pressure},
and metal-insulator transition~\cite{fazekas1980charge, wang2019magic}.
Indeed, the NFL behavior with the resistivity exponent varying with pressure or gate voltage was observed in 1T-TiSe$_2$
\cite{li2016controlling,kusmartseva2009pressure}.
 
Motivated by these observations, we will study the electric conduction through the network superstructure.
The electronic properties of conducting networks have been studied theoretically in various contexts. For instance, our previous works~\cite{Park-Prep, lee2020stable} have shown that 1T-TaS${}_2$ in nearly-commensurate charge-density wave states hosts a conducting honeycomb network via STM~\cite{Park-Prep} and that the network supports a cascade of \textit{anomalously stable} flat bands, which can explain unusual enhancement of the superconductivity~\cite{lee2020stable}, and higher-order topology~\cite{lee2020stable, PhysRevMaterials.3.114201}. Also the network systems have received some attention in connections with the phenomenology of the magic-angle graphene and Chalker-Coddington physics~\cite{wu2019coupled, PhysRevB.100.115128, cao2018correlated,cao2018unconventional,efimkin2018helical,chalker1988percolation,chou2020hofstadter,PhysRevLett.125.096402,PhysRevLett.123.137202,Vitor-NFL}. However, systematic investigation of electric conduction through the network in the presence of electron interaction has been largely lacking (see also phenomenological discussions in \cite{Vitor-NFL, medina2013networks}). In this Letter, we take a first step toward elucidating universal NFL behaviors in networks.

First, generalizing Landauer-B\"{u}ttiker approach, we can naturally derive the macroscopic Pouillet's law so that the conduction of the entire network is characterized by the conductivity of a single junction.
Furthermore, by including the effects of the electron-electron interactions, we find a remarkably broad set of NFL behaviors emerging naturally in the conducting network systems. This originates from the Tomonaga-Luttinger liquid (TLL) nature of the 1D segments of interacting electrons. We will explain when and why NFL behaviors are expected. Furthermore, we will argue that these NFLs can be \textit{one-to-one} matched with the characteristics of the junctions, i.e., the ``boundary conditions" for electrons at the junctions. As a consequence, the resistivity exponent $\alpha$ of the network is determined by the Luttinger parameter $K$ which describes each 1D segment, potentially explaining the intriguing variation of the resistivity exponent observed in the experiments~\cite{li2016controlling,kusmartseva2009pressure}.

\begin{figure}[t]
\centering{ \includegraphics[width=1.0\columnwidth]{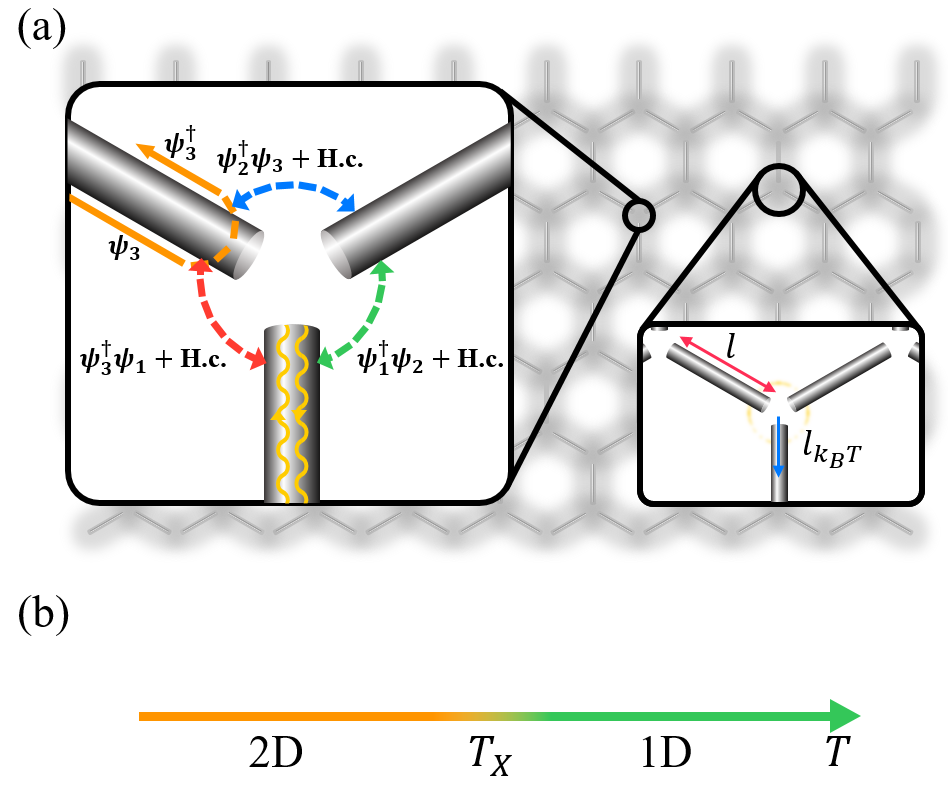} }
\caption{
(a) Schematic picture of honeycomb network. Each link is a TLL Eq.\eqref{LL}. The left inset shows a possible $H_{Y; (a,b,c)}$ around a single junction. The right small inset represents the scales of the problem in our RG process. Here $l_{k_{B} T}$ is the thermal coherence length at temperature $T$ and $l_{k_{B} T}<l$ for $T > T_X$ where $T_X$ is the crossover temperature. (b) ``Phase diagram" of the network in temperature.
}\label{Fig1}
\end{figure}

{\color{green} \textbf{\textit{2. Model}:}}
In this Letter, as an example, we will mainly consider the minimal honeycomb network model, which consists of 1D segments of interacting spinless electrons as in [Fig.\ref{Fig1}(a)].
The network is analyzed in terms of Renormalization Group (RG), starting from the microscopic energy scale (such as the bandwidth $W$ of the 1D wire). The RG transformation should be terminated at the energy scale given by the temperature $T$.
At sufficiently low temperatures $T \ll W$, the interacting electrons in the 1D segment of length $l$ can be described as a TLL characterized by a Luttinger parameter $K$ and the (renormalized) velocity $v_F$:
\begin{align}
H_{TLL; a} = \frac{v_F}{2} \int^{l}_{0} dx~ \Big[K(\partial_x \phi_a)^2 + \frac{1}{K} (\partial_x \theta_a)^2 \Big] ,
\label{LL}
\end{align}
where $a$ is the index labelling each 1D segment. Here $v_F$ and $K$ are different from those of the bare, free electrons due to interactions between them.
See [\onlinecite{suppl_ref}] for our bosonization convention.
Here we assume that the electron filling per wire is incommensurate to avoid unnecessary complications and that there is no disorder.
The effective Hamiltonian of the whole network reads
\begin{align}
H = \sum_{a} H_{TLL; a} + \sum_{\langle a,b,c\rangle} H_{Y; (a,b,c)},
 \label{Network}
\end{align}
where $H_{Y;(a,b,c)}$ is the \textit{local} interactions between the three neighboring wires around the Y-junctions, such as the hopping between the wires. See [Fig.\ref{Fig1}(a)]. Although the precise form of $H_{Y;(a,b,c)}$ depends on microscopic details, we will show that the essential NFL behaviors are independent of them and thus universal.
In addition, we will assume coupling of electrons with external environment (typically phonons). However, the result is again independent of its precise form as we will explain later.

Strictly speaking, our model is based on the standard TLL theory, which applies to systems with only short-range interactions. Nevertheless, the long-range Coulomb interaction between electrons is often screened. Indeed, the TLL behaviors are rather ubiquitous in actual quasi-1D materials, e.g. \cite{Bockrath1999, Ma2017, Kim2007, Zhen1999, Zhao2018}. 
As for the candidate materials hosting network superstructures, the layered quasi-2D systems, TaS${}_2$ and TiSe${}_2$, are metallic~\cite{li2016controlling,sipos2008mott} and thus the screening is expected. For 2D materials such as twisted bilayer graphene, substrates~\cite{Kim2020} can provide screening of long-range Coulomb interactions.
Therefore our model will describe a wide variety of material realizations of electrons on networks.


{\color{green} \textbf{\textit{3. Dimensional Crossover}:}}
The energy scale $\epsilon$ can be translated to the lengthscale $l_\epsilon \sim \hbar v_F / \epsilon$.
Thus we can introduce the crossover temperature
\begin{align}
T_{\text{X}} \sim  \frac{\hbar v_F}{k_B l} , \label{Tx}
\end{align}
which is the temperature where the thermal coherence length touches the wire length $l$.\cite{kane1992transmission}
For $T \lesssim T_{\text{X}}$, the electrons can `feel' the finiteness of the wire length and recognize the network as the 2D system. The physics of this regime is essentially two-dimensional.
On the other hand, above the crossover temperature, $T \gtrsim T_X$, we are probing the system at the lengthscale shorter than the segment length $l$. Thus the system does not ``know'' that the wires form a 2D network, and the physics is largely governed by the properties of 1D wires and 0D junctions.
Based on these observations, we can draw a ``phase diagram" in $T$ [Fig.\ref{Fig1}(b)].
In real materials, we roughly estimate $T_{\text{X}} \gtrsim \mathcal{O} (90)$K in 1T-TaS${}_2$ and $T_{\text{X}} \gtrsim O(50)$K in twisted bilayer graphenes (for $l = 140$ nm, though it is tunable). For 1T-TiSe${}_2$, we estimate $T_X \gtrsim \mathcal{O} (6)$K with some assumptions. For the details and possible implications in existing experimental data, see [\onlinecite{suppl_ref}].

 We compare our crossover temperature Eq.\eqref{Tx} of the network with that of coupled wire systems or sliding LLs\cite{Wen1990,SCHULZ1991, Castellani1992,Brazovskii1985,Bourbonnais1986,giamarchi2003quantum, Guinea1993,Sondhi2001,Mukhopadhyay2001,Emery2000,Vishwanath2001}. There, the wires are aligned along the same directions, and we assume only the short-ranged terms. Then the crossover temperature toward a regular 2D FL from the 1D TLL limit depends on both the intra- and interwire terms~\cite{Wen1990,SCHULZ1991, Castellani1992,Brazovskii1985,Bourbonnais1986,giamarchi2003quantum, Guinea1993}
\begin{align}
T_{X}^{\text{coupled wires}} \sim t_\perp \cdot \left(\frac{t_\perp}{W}\right)^{\frac{1}{2-\Delta(K)}}, \label{Tx-Stripe}
\end{align}
where $t_\perp$ is the strength of the interwire hopping, and $\Delta(K)$ is its scaling dimension. 




 
In this Letter, we focus on the electric conduction in the network in the ``1D TLL physics'' regime
\begin{align}
  T_X \lesssim T \ll W ,
  \label{1Dregime}
\end{align}
which has not been much explored.
Since each segment is described by a scale-invariant TLL, the temperature dependence of the transport can be associated to the properties of the junctions. Hence, to study the network in this regime, we can utilize the RG analysis of junctions of TLLs \cite{kane1992transport,kane1992transmission,furusakinagaosa,chamon2003junctions,oshikawa2006junctions}. Indeed, we find that the emergent NFLs in the network can be characterized by the RG fixed points of a single junction of TLLs. We will show that the RG fixed points of the junctions determine not only the leading dc conductivities but also leading scaling corrections, which are power-law in temperature $T$.
Namely, the NFL behavior is universal in the 1D regime~\eqref{1Dregime} where the junction is described by a RG fixed point, which pins down the scaling dimensions of all the possible perturbations.
Below we will explicitly illustrate this for the two simplest fixed points~\cite{oshikawa2006junctions, chamon2003junctions}, namely ``disconnected fixed point" ($K>1$) and ``connected fixed points". As our terminology suggests, the connected fixed point gives rise the maximum conductances between the wires~\cite{oshikawa2006junctions}. Some properties and stability of these fixed points are reviewed in \cite{suppl_ref}.

\begin{figure}[t]
\centering{ \includegraphics[width=1.0\columnwidth]{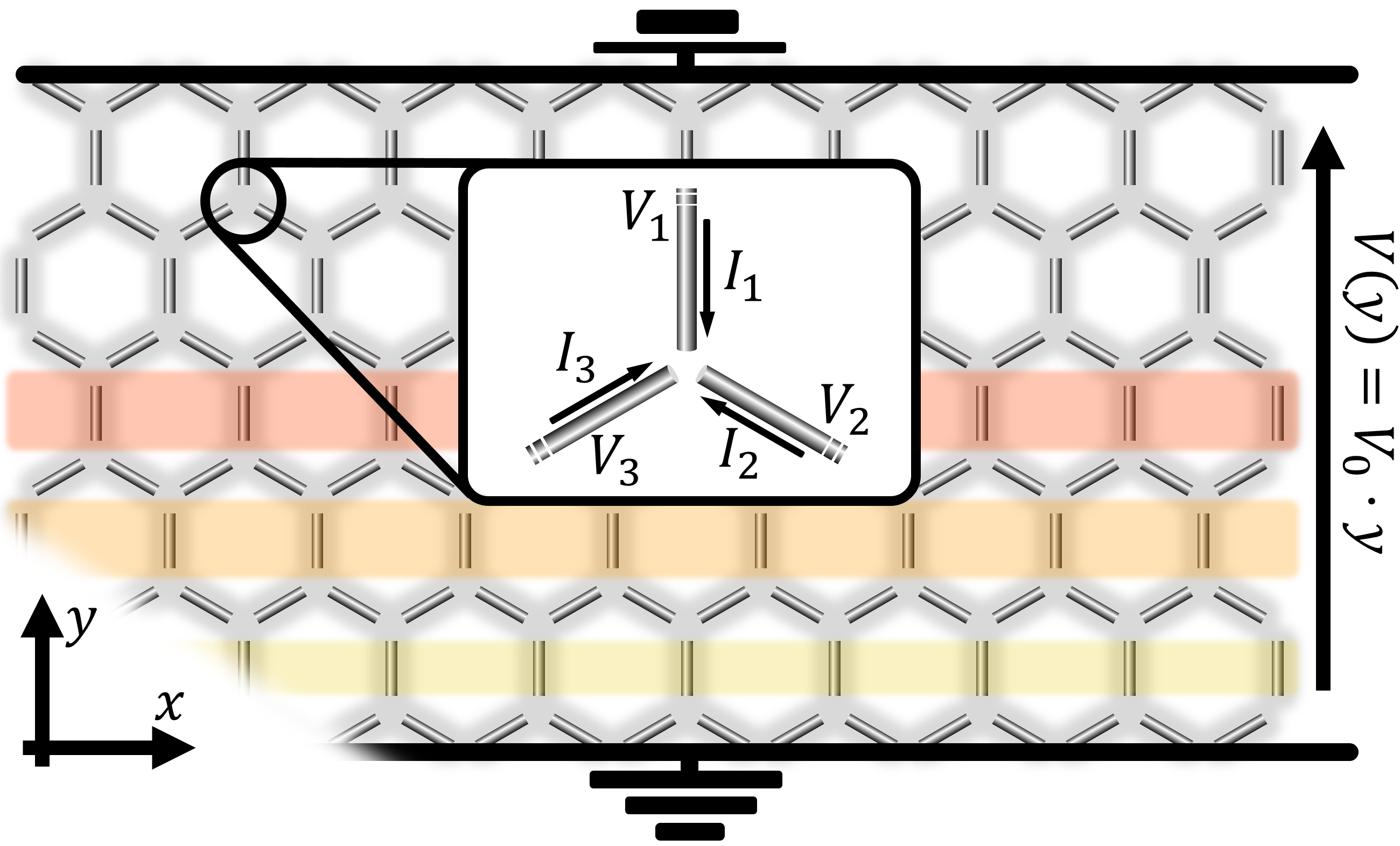} }
\caption{ Conduction through the network. The inset represents a single Y-junction, where voltage and current satisfy $I_a = \sum_b G_{ab} V_b$. The voltage is uniform in $x$ and increases in $y$, as depicted in the figure. Wires with the regions of the same colors have the same voltage. 
}\label{Fig2}
\end{figure}

{\color{green} \textbf{\textit{4. Conduction through the Network}:}}
Here we show that the resistivity of the network is in general given by the power law~\eqref{eq.resistivity_power}
where the exponent $\alpha$ is determined by the Luttinger parameter of the 1D TLL segment and the boundary conditions at the junctions. 
 More precisely, $\alpha = 2\Delta(K) - 2$ where $\Delta(K)$ is the scaling dimension of the leading irrelevant operators at the junctions. The numerical coefficient $c$ depends on microscopic details. For example, for the disconnected fixed point, $\rho_0 = 0$ and
\begin{align}
  \alpha(K) = 2K - 2 .
  \label{eq.alpha_N}
\end{align}
 To establish this, we first show that the 0D electric \textit{conductance} at a single junction determines the 2D \textit{conductivity} of the whole network. For this, we imagine a perfect 2D network made out of the identical Y-junctions [Fig.\ref{Fig1}(a)]. We apply a uniform voltage drop across the $y$-direction and then calculate the electric current flowing through the network [Fig.\ref{Fig2}].
In materials, electrons interact with the external environment such as phonons. In this Letter, as we consider the fairly high temperatures $T \gtrsim T_X$, we assume that the electron-phonon coherence length is shorter than the segment length $l$. Then the electrons in the segment equilibrate, so that each segment has a well-defined voltage.
Under this assumption, we can immediately compute the conductivity of the whole 2D network out of the ``conductance tensor'' of a single Y-junction.

Each fixed point of the Y-junction can be characterized by its $3 \times 3$ conductance tensor $G_{ab}$, which relates the electric current at the $a$-th wire with the voltage at the $b$-th wire [Fig.\ref{Fig2}], i.e., $I_a = \sum_b G_{ab} V_b$.
For the spinless fermions, all the known fixed points~\cite{oshikawa2006junctions, chamon2003junctions} respect the $\mathbb{Z}_3$ permutation symmetry between the three neighboring wires. Hence the conductance tensor can be parameterized only by the two numbers $G_{S}$ and $G_A$.  Imposing $I_a = \sum_b G_{ab} V_b$ at all the junctions, we can fix the electric current at every wires. From the current, we obtain the conductance of the entire network,\cite{suppl_ref} which is found to be proportional to the width, and inversely proportional to the length. (Similar discussions were given in \cite{Amato-Pastawki_PRB1990, Buttiker-decoherence_PRB1986}.) 
In this way, our network construction leads naturally to the classical Pouillet's law, and the conduction property of the system can be characterized by constant \textit{conductivity} tensors  
\begin{align}
g_{xx}(T) = \sqrt{3} G_S(T)/4, \quad g_{xy} (T) = G_{A}(T)/4.
\label{eq.conductivity}
\end{align}
The factors of $1/4$ and $\sqrt{3}/4$ have the geometric origin, e.g., the size of the unitcell. We also checked $g_{xx}(T)=g_{yy}(T)$ and $g_{xy} (T) = - g_{yx}(T)$ \cite{suppl_ref}. While the classical nature of the conduction is a natural consequence of the assumed local thermalization (and thus decoherence) in each segment, it is remarkable that the macroscopic property (conductivity) is determined by the property of the microscopic junctions, independently of the details of the thermalization. This can be regarded as a generalization of Landauer-B\"{u}ttiker approach~\cite{Datta-book} to extensive macroscopic systems of interacting electrons.


Hence, we next compute $G_S$ and $G_A$ of an Y-junction at finite temperatures $T$. For this, we generalize the results of \cite{oshikawa2006junctions, chamon2003junctions, kane1992transmission,kane1992transport} to the Y-junctions.
For instance, at the disconnected fixed point, the leading perturbation is the interwire hopping [Fig.\ref{Fig1}(a)]
\begin{align}
H_{Y} = -t \sum_{j=1,2,3} \psi^{\dagger}_{j}(0)\psi_{j-1}(0) + H.c., 
\nonumber
\end{align}
where $\psi_0$ is identified with $\psi_3$, and $\psi_{j}(0)$ represents
the electron annihilation operator of the $j$-th wire at the Y-junction, i.e., $x=0$. 
Up to the 2nd order in $t$, we find~[\onlinecite{suppl_ref}]
\begin{align}
G_{S} (T) \approx 0 + \frac{2 e^2 t^2}{h} \left(\pi \tau_c\right)^{2K} \frac{\Gamma\left(\frac{1}{2}\right)\Gamma\left(K\right)}{\Gamma\left(\frac{1}{2}-K\right)} \cdot T^{2K-2}, \label{Cond-Neumann}
\end{align}
with $\tau_c$ being the inverse of an UV cut-off $W$ of Eq.\eqref{LL}. The first term $0$ in $G_S$ is the universal conductance of the disconnected fixed point.\cite{oshikawa2006junctions, chamon2003junctions} The second term $\sim T^{2K-2}$ represents the correction from the leading irrelevant perturbation $H_{Y}$ above. They are evaluated within the standard linear response theory combined with the perturbative expansion in $H_{Y}$\cite{kane1992transmission,kane1992transport,suppl_ref}. The correction $\sim T^{2K-2}$ is consistent with Eq.\eqref{eq.resistivity_power}, because the scaling dimension of $H_Y$ is $\Delta(K)=K$.
As we have seen in Eq.\eqref{eq.conductivity}, this conductance of the single junction directly gives the conductivity of the 2D network and the resistivity exponent is given as in Eq.\eqref{eq.alpha_N}.  A fixed point is stable as far as the scaling dimension of the perturbation is larger than 1. Hence the disconnected fixed point is stable for $\Delta(K) = K >1$. Furthermore, this disconnected fixed point is the only known stable fixed point for repulsive interactions\cite{oshikawa2006junctions, chamon2003junctions} and so this ``strange insulator'' behavior with the power-law divergence of the resistivity at low temperatures is generic.  

For attractive interaction, $K<1$, the inter-wire tunneling is a relevant perturbation.
Eq.\eqref{Cond-Neumann} could describe the conductivity at higher temperatures if the inter-wire coupling at the junction in the microscopic model is weak. In this case, the same expression~\eqref{Cond-Neumann} now describes the power-law with a positive resistivity exponent~\eqref{eq.alpha_N}, namely the typical behavior of a \emph{metallic} NFL.
When the attractive interaction is sufficiently strong so that $K<1/3$, as the temperature is lowered, the junction is governed by the connected fixed point. This fixed point is stable for $K<1/3$ because the scaling dimension of the leading irrelevant operators at this fixed point is $1/3K$\cite{oshikawa2006junctions,suppl_ref}. Taking the operators into account, we can again evaluate the conductance within the standard linear response theory\cite{kane1992transmission,kane1992transport,suppl_ref}
\begin{align}
G_{S} (T) \approx \frac{2}{3K} \frac{e^2}{h} - C \cdot T^{\frac{2}{3K}-2},
\label{Cond-Dirichlet}
\end{align}
where we suppressed all the unimportant constants into $C$. As before, the first term is the universal conductance of the connected fixed point,\cite{oshikawa2006junctions, chamon2003junctions} and the second term  $\sim T^{2/(3K)-2}$ is the correction from the perturbative expansion of the leading irrelevant operators.\cite{suppl_ref}  If the inter-wire coupling at the junction is strong, Eq.\eqref{Cond-Dirichlet} would describe the entire temperature range where the 1D TLL description is valid. This also gives the metallic NFL behavior with the decreasing resistivity at lower temperatures, but with the resistivity exponent $\alpha = 2/(3K)-2 >0$.  

For both the disconnected and connected fixed points, $G_A$ and thus the Hall conductivity $g_{xy}$ vanish, as expected from the time-reversal invariance of the underlying model. On the other hand, for the network of $1/3<K<1$ under a uniform magnetic field, the chiral fixed point of $G_A \neq 0$ is stabilized:
\begin{align}
G_{S} (T) \approx \frac{4K}{1+3K^2} \frac{e^2}{h}, ~G_{A} (T) \approx \frac{4}{1+3K^2} \frac{e^2}{h}, \label{Cond-Chiral}
\end{align}
up to the perturbative correction $\sim \mathcal{O}\left(T^{8K/(1+3K^2)-2}\right)$. Such network is metallic and has the ``universal'' Hall conductivity, which is determined by the Luttinger parameter~\cite{medina2013networks, suppl_ref}. This result is consistent with \cite{medina2013networks}, which considered the electric conduction across a network consist of the chiral fixed point. We note that however, \cite{medina2013networks} missed much of the NFL physics which we explored here. See the comparison in \cite{suppl_ref}. While the realization of the chiral fixed point requires an explicit breaking of the time-reversal invariance, in principle the required breaking can be infinitesimal, e.g., by a very weak magnetic field through the junctions~\cite{chamon2003junctions,oshikawa2006junctions}. This behavior is again quite different from a normal metal, in which the Hall conductivity is proportional to the applied magnetic field. We also note that for $K>1$ or $K<1/3$, the chiral fixed point is unstable. Instead, the connected and disconnected fixed points are stable as discussed above.

Let us give a few remarks on Eqs.\eqref{Cond-Neumann},~\eqref{Cond-Dirichlet}, and~\eqref{Cond-Chiral}, which represent the conductance in the vicinity of three different RG fixed points for the junction. First of all, in all cases, the conductivity of the network exhibits a power law in temperature, whose exponent continuously evolves as the Luttinger parameter $K$ varies. This is the manifestation of the exact marginality of the Luttinger parameter. Essentially, this marginality allows the scaling dimension of electrons vary smoothly, which translates as the continuously changing $\alpha$ in Eq.\eqref{eq.alpha_N}. In experiment, this means that, as the external parameters, e.g., pressure and gating, are tuned, the exponent of the temperature dependence of the resistivity will continuously change. This is markedly different from the behavior of a regular Fermi Liquid, where the exactly marginal deformation, i.e., the change of the Fermi velocity, does not alter the temperature dependence of the transport coefficients.

Finally, we comment on possible effects of disorders on transport. There are distinct types of the disorders at different length scales. For instance, microscopic impurities in TLLs will induce a power-law correction $\sim T^{2/K-2}$ to the conductivity,\cite{kane1992transmission,kane1992transport, Giamarchi1996} which will add up to those from the junctions. Similarly, randomly missing (completely disconnected) Y-junctions
will introduce an additional $\sim T^{2K-2}$ correction. Details are given in \cite{suppl_ref}.

{\color{green} \textbf{\textit{5. Conclusions}:}} We have demonstrated the emergence of a novel class NFLs in the conducting networks, whose universal properties are controlled by the RG fixed points of the junction of TLLs.
This makes our network system a unique theoretical platform, where the isotropic NFL behaviors in $d \geq 2$ dimensions can be deduced from the well-established theory on strongly correlated electrons in 1D.
The NFLs we have proposed are potentially already out there\cite{xu2019giant, sipos2008mott, li2016controlling, kusmartseva2009pressure, suppl_ref} in experiments, and/or can be easily realized and verified in currently-available setups. For instance, one can artificially pattern the network superstructure in experiments\cite{forsythe2018band}. In twisted bilayer graphenes, the crossover temperature $T_X$, which is related with the length $l$ of the underlying wires, can be controlled by tuning the twisting angle. Hence, in these systems one can look at the dependence of the conductivities on temperature and twisting angles to observe the emergence of the putative NFLs, which will be quite spectacular!

Once the power-law behavior of the resistivity is observed, the Luttinger parameter $K$ for the TLL describing each segment is inferred from the resistivity exponent $\alpha$. Our scenario can then be verified by a consistency check with an independent determination of the Luttinger parameter of the 1D segment, for example by ARPES measurement of the local density of states $\sim |\omega|^{(K+1/K)/2 -1}$~\cite{ishii2003direct}.  The Luttinger parameter and scaling dimensions $\Delta(K)$ of the leading irrelevant operators at the junctions can be also extracted from the specific heat $C_v(T)$ and susceptibility $\chi (T)$. For example, the specific heat has the two contributions, one from the 1D TLLs and the other from the junctions.\cite{Fujimoto2003,Fujimoto2004,Furusaki2004} The 1D part scales as $\sim T$, but the junction contribution has $\sim T^{2\Delta(K) - 2}$.\cite{Fujimoto2003,Fujimoto2004,Furusaki2004} Hence $C_v \sim c_1 T + c_2 T^{2\Delta(K) - 2}$ in total and similarly $\chi \sim b_1 + b_2 T^{2\Delta(K)-3}$.
In the future, it will be interesting to investigate explicitly the effect of the long-range Coulomb interaction on the network NFLs,
following the related studies on a single TLL and coupled wires~\cite{Kane1997, Wang2001, Vu2020,Schulz1993,Sur2017,Kopietz1995, Kopietz1997}.

\acknowledgements
We thank Chenhua Geng, Jung Hoon Han, Jun-Sung Kim, Gil-Ho Lee, Sung-Sik Lee, Jeffrey Teo and Han Woong Yeom for helpful discussion. JL and GYC are supported by the National Research Foundation of Korea (NRF) grant funded by the Korea government(MSIT) (No. 2020R1C1C1006048 and No.2020R1A4A3079707) and also by IBS-R014-D1. This work is supported by the Air Force Office of Scientific Research under award number FA2386-20-1-4029.  MO is supported in part by MEXT/JSPS KAKENHI Grants  No. JP18H03686 and No. JP17H06462, and JST CREST Grant No. JPMJCR19T2.  We also thank Claudio Chamon and Dmitry Green for bringing our attention to the reference \cite{medina2013networks} and helpful discussions.

\bibliographystyle{apsrev}
\bibliography{Network}

\end{document}


\title{Supplemental Information for ``Non-Fermi Liquids in Conducting $2D$ Networks"}

\author{Jongjun M. Lee}
\thanks{Electronic Address: michaelj.lee@postech.ac.kr}
\affiliation{Department of Physics, Pohang University of Science and Technology (POSTECH), Pohang 37673, Republic of Korea}

\author{Masaki Oshikawa}
\thanks{Electronic Address: oshikawa@issp.u-tokyo.ac.jp}
\affiliation{Institute for Solid State Physics, The University of Tokyo, Kashiwa 277-8581, Japan}
\affiliation{Kavli Institute for the Physics and Mathematics of the Universe, Kashiwa 277-8583, Japan }
\affiliation{Trans-scale Quantum Science Institute, University of Tokyo, Bunkyo-ku, Tokyo 113-0033, Japan}

\author{Gil Young Cho}
\thanks{Electronic Address: gilyoungcho@postech.ac.kr}
\affiliation{Department of Physics, Pohang University of Science and Technology (POSTECH), Pohang 37673, Republic of Korea}
\affiliation{Center for Artificial Low Dimensional Electronic Systems,  Institute for Basic Science (IBS),  Pohang 37673, Korea}

\date{\today}
\maketitle
\tableofcontents
\appendix
\section{Bosonization Convention}
The standard Tomonaga-Luttinger Liquid (TLL) theory in the main text is based on the bosonization:
\begin{equation}
\begin{aligned}
S_0 &= \int d\tau dx \frac{u}{2}\Big(K(\partial_{x}\phi(x,\tau ))^2 +\frac{1}{K}(\partial_{\tau}\theta(x,\tau))^2\Big).
\end{aligned}
\end{equation}
The electron operators are bosonized as 
$
\psi_{L} \sim e^{-i\sqrt{\pi}(\phi(x)+\theta(x))},  \psi_{R} \sim e^{+i\sqrt{\pi}(\phi(x)-\theta(x))}.
$
Hence $\phi(x)$ represents the phase of the charge-density wave order, and $\theta(x)$ is the phase of the superconductivity. For instance, the superconducting order parameter is given by 
$
:\psi_R (x) \psi_L (x) : \sim \exp\left[ -2i\sqrt{\pi} \theta (x) \right]
$.
On the other hand, the charge-density wave can be written as 
$
:\psi^{\dagger}_R(x) \psi_L (x) : \sim  \exp\left[ 2i\sqrt{\pi} \phi (x) \right]
$. 
The normal ordering $: O :$ will be implicit. Hence it will never appear explicitly in both the main text and supplemental materials. Note that, in this representation, $K>1$ ($K<1$) is the case with the repulsive (attractive) interactions between electrons. 


\section{Experimental Realizations of Networks \& $T_X$}
Here we review some details of experimental realizations of networks (which are mentioned in the main text) and the assumptions in calculating $T_X$ in those materials. 

\subsection{1T-TaS${}_2$} 
The network appears in the nearly-commensurate charge-density wave phase, which can be accessed by disordering the commensurate charge-density waves in a various ways. The detailed STM analysis of the network is given in our previous works. In this materials, the conducting network is formed via the current-carrying domain walls of the length $\sim O(70-80)$ \si{\angstrom}, forming a regular honeycomb array.\cite{Park-Prep} The first-principle DFT calculation on the domain wall electronic structure has been carried out and the Fermi velocity was roughly estimated to be $v_F \sim 8.5 \times 10^4$ m/s.\cite{Park-Prep} Putting all the factors into $T_X$, we find  
\begin{align}
T_X \sim \frac{\Big(6.6 \times 10^{-34} ~\text{[J $\cdot$ s]} \Big)\cdot \Big(8.5 \times 10^4 ~\text{[m/s]}\Big)/2\pi}{\Big(1.38 \times 10^{-23} ~\text{[m${}^2$ kg s${}^{-2}$ K${}^{-1}$]}\Big) \cdot \Big(70 \times 10^{-10} ~\text{[m]} \Big)} \sim 92 K. 
\end{align}
Experimentally, when the sample is pressurized, the power of the resistivity in temperature is unusual\cite{sipos2008mott} though the precise exponent is unclear. To make the connection more concrete with the non-Fermi liquids in our main text, more detailed analysis will be necessary. Finally, we note that this $T_X$ is obtained under the assumption that the network structure is rigid under lowering the temperature and changing the parameters in experiments. 

\subsection{Twisted Bilayer Graphene} 
When the two layers of graphene are twisted each other for small angles, then the helical network appears. For instance, for the experimentally available sample, the length scale is roughly of $\sim O(140)$ nm at the twisting angle 0.1${}^{\circ}$.\cite{xu2019giant} We assume that the electronic motion is one-dimensional. The Fermi velocity is taken roughly as $10^6$ m/s. Then, with this assumption, we can immediately obtain $T_X \approx 54 K$, which is well below the bulk gap inside AB and BA domain regions $\sim 50$ meV.\cite{xu2019giant} Note that there is an insulator-like transport $\rho_{xx}(T)$ for $T>100$K, which is above $T_X$ but well below the energy gap AB/BA domain regions. The authors attributed this to the thermally-activated transport across the domains (instead of the domain walls). This behavior is reminiscent of the ``strange insulator" behaviors in our theory, though more serious investigation would be desirable. In any case, this twisted bilayer graphene system is famous for possibly engineering of the length scale of the network, whose moiré superstructure unitcell is roughly scaling as $L \sim a_0/\theta$, in which $\theta$ is the twisting angle between the layers and $a_0$ is the size of the atomic unitcell of a graphene.  

\subsection{1T-TiSe${}_2$}
This material supports charge-density wave orders. There are several ways to destroy the commensurate charge order and turn them into the nearly-commensurate or incommensurate ones. For example, one can apply the gating to the sample.\cite{li2016controlling} In the system under the gating, the regular oscillations in resistivity due to the magnetic field was observed in the superconducting phase, i.e. so-called Little-Parks effect.\cite{li2016controlling} This implies that the superconductivity in the system is textured and that the low-energy electrons participating to the superconductivity also flow along the network. When combined with the Landau-Ginzburg theory, one can actually derive the length scale of the network, which was estimated to be roughly $\sim O(70)$ nm.\cite{chen2019discommensuration} Assuming that the low-energy electrons are moving as in 1D systems and their Fermi velocity is roughly the same as its bulk excitations $\sim 6.1 \times 10^4$ m/s,\cite{qian2007emergence} one can estimate $T_X \approx 6 K$. Note that the STM study on the surface of Cu-intercalated TiSe${}_2$ found that the electronic density of states are indeed enhanced at the domain walls.\cite{PhysRevLett.118.106405} 

However, there are two points to be careful about 1T-TiSe${}_2$. First, in contrast to 1T-TaS${}_2$, domain regions of the charge-density wave are not insulating. Second, the Landau-Ginzburg theory seems to suggest that the domain walls are rather fat,\cite{chen2019discommensuration} though the STM study finds that the domain wall width is very narrow.\cite{PhysRevLett.118.106405} Both of these effects may weaken our assumption that the low-energy electrons in 1T-TiSe${}_2$ is one-dimensional. In any rate, in 1T-TiSe${}_2$, non-Fermi transports were reported in both the gated and pressurized ones.\cite{li2016controlling, kusmartseva2009pressure}   

\subsection{Ultracold Gas Systems}
To our knowedlge, the large-scale network of 1D metals has not been realized in ultracold gas experiment yet. However, we would like to note that the ultracold gas systems have a potential to realize the network of the 1D quantum metals and their junctions. First, there has been some experimental realizations of TLLs in optical trap systems\cite{Paredes2004, Haller2010, Angelakis2011}. Second, there is a tool that can apply narrow constriction potentials to atoms. It is called as the Digital Mirror Device (DMD), which is an array of programmable microscale mirrors. This tool has been already implemented and explored in several groups.\cite{Kwon2020,Fukuhara2013, Hausler2017, Mazurenko2017, Schafer2020} For instance, one group recently simulated not only one dimensional fermions, but they also experimented its transport property by constructing a tunneling junction \cite{Kwon2020}. The DMD, which can apply various shapes of potentials to the ultracold gas, can potentially help to realize a Y-junction in the ultracold gas systems, which will serve as a building block of the network.

\section{2D Conductivity \& Comparison with Medina et al.}
Here we present some details of calculation of 2d conductivity of the \textit{isotropic} network from the ``conductance" of a single Y-junction and also the comparison of our work with Medina \textit{et. al.} [\onlinecite{medina2013networks}]. Note that, although we focus on the honeycomb newtork here, it is straightfoward to generalize our approach here to any shape of the networks, e.g. square and triangular networks.  

\begin{figure}[ht]
\includegraphics[scale=0.6]{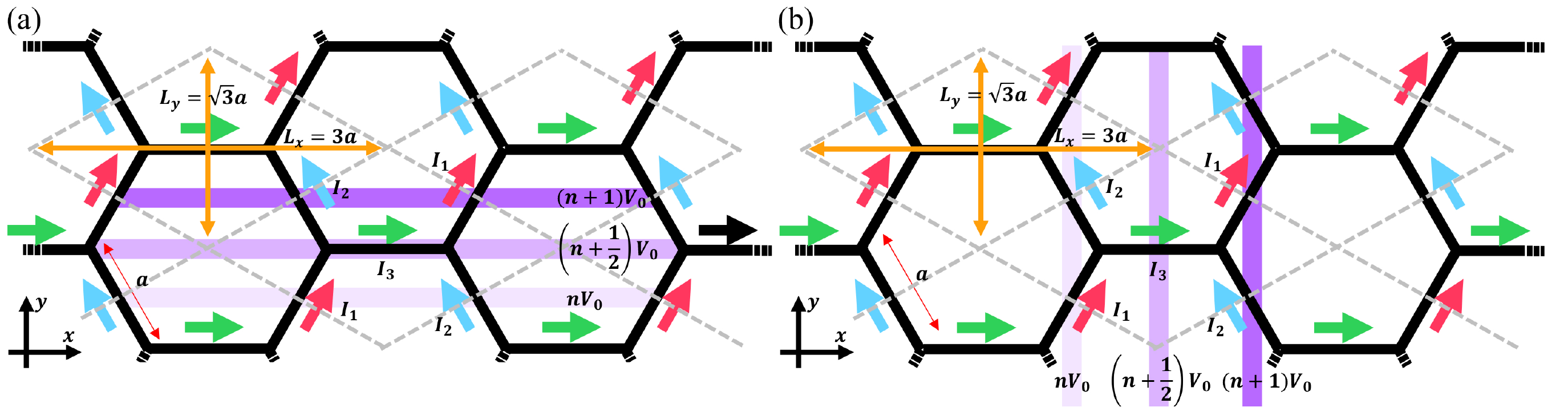}
\caption{Honeycomb Networks. (a) The voltage is applied from the top to the bottom along the $y$-axis as much as $V_0$ per an unitcell. The periodic boundary condition was assumed along the $x$-direction. (b) The voltage is applied from the right to the left along the $x$-axis as much as $V_0$ per unitcell. The periodic boundary condition was assumed along the $y$-direction. The arrows indicate the direction of the current and the colored lines represent the equipotential lines. The gray dotted parallelogram represents the unitcell of the network.  The length $(L_x =3a,\: L_y =\sqrt{3}a)$ for the unitcell is also written in the Figures. }
\label{General_Array}
\end{figure}

- \textbf{2D Conductivity}: The honeycomb network is a regular 2D array of Y-junctions. See Fig. \ref{General_Array}. We will use the formula for the conductance tensor at each junction, i.e. $G_{ij}=G_S (3\delta_{ij}-1)/2 +G_A \epsilon_{ij}/2$. From this, we  calculate the 2D conductivity in the two orientations in Fig.\ref{General_Array} and compare them. 

To begin with, let us consider a case where the voltage drop is applied along the $y$-direction [Fig. \ref{General_Array}(a)]. The total voltage drop across the sample is  $NV_0$ where $N$ is the number of unitcells in the $y$-direction. Hence the voltage drop per unitcell is $V_0$. Utilizing the translation symmetry and periodic boundary condition along the $x$-axis, we can calculate the 2D conductivity from a single unitcell. See the gray dotted parallelogram in [Fig. \ref{General_Array}(a)]. We find 
$
I_1 = -\frac{G_A +3G_S}{4} V_0 , ~
I_2 = \frac{G_A -3G_S}{4} V_0 , ~
I_3 = -\frac{G_A }{2}V_0
$ by using $I_j = \sum_{i}G_{ji}V_i$. What we need to do next is to calculate the longitudinal and Hall currents per unitcell are
$
I_y = I_1 +I_2 = -\frac{3G_S V_0}{2} , I_x = I_3 = -\frac{G_A}{2}V_0
$.

Finally, to calculate the conductivity, we need to obtain the current densities by dividing the currents $\{ I_x, I_y \}$ by proper length scales, i.e., the size of the unitcell. See the size of the gray dotted parallelogram in the Fig. \ref{General_Array}. Hence, we find 
\begin{equation}
\begin{aligned}
J_x &= \frac{I_x}{L_y} = -\frac{G_A V_0}{2\sqrt{3}a},\; J_y = \frac{I_y}{L_x} = -\frac{G_S V_0}{2a}.
\end{aligned}
\end{equation}
Here $L_y = \sqrt{3}a$ and $L_x = 3a$. Now, one uses $J_i = g_{ij}E^j$ to compute the conductivity, where $\vec{E} = -2V_0/L_y \hat{y}$. Therefore, we obtain the two components of the 2D conductivity tensor $g_{yy} = \frac{\sqrt{3}G_S}{4}, g_{xy} = \frac{G_A}{4}$. Similarly, we can consider the case that the voltage drop is applied along the x-direction [Fig. \ref{General_Array}(b)]. The total voltage drop across the sample is $NV_0$ where $N$ is the number of unitcells in the $x$-direction. From this, we find $g_{xx} = \frac{\sqrt{3}}{4} G_S , g_{yx} = -\frac{G_A}{4}$. Hence we find $g_{xx}=g_{yy}$ and $g_{xy}=-g_{yx}$, i.e. the 2D conductivity is rotationally symmetric (as expected).

- \textbf{Comparison with Ref}.[\onlinecite{medina2013networks}]: Note that the resistivity can be also calculated by taking the inverse of the conductivity. In Ref. \onlinecite{medina2013networks}, the resistivity tensor specifically for the network consist of the chiral fixed point is calculated. For this particular fixed point, if we invert our conductivity tensor, then we find that our result agrees with Ref. \onlinecite{medina2013networks}. 

We woud like to comment on a few perspectives of our work, which go beyond the reference\cite{medina2013networks}. First, we have clarified that the non-Fermi liquid (NFL) is generically expected in the network due to the underlying 1D TLL. Second, we have elucidated the relation of the NFL physics with the properties of the boundary fixed points, which can thus classify various NFLs emerging in the network. Thirdly, we have explained why the NFL phenomena is universal against e.g. the perturbations. Fourthly, we also figured out in which temperature region such NFL behavior is actually expected. Furthermore, we have explicitly computed the NFL transport for the connected and disconnected fixed points, which were not discussed in \onlinecite{medina2013networks}. Given that many experiments on real network materials were carried out in the absence of the magnetic field, these fixed points are likely more important in realistic setups than the chiral one. Finally, we have shown that the most prominent characteristics of the NFL is the smoothly-varying temperature exponent in transport, which is related with the Luttinger parameter and the dimension of leading irrelevant operators. We also have explored other experimentally important properties such as disorder, structural deformation, specific heat, susceptibility, and spectral functions, which could be useful in confirming our scenario in experiments. 


\section{Properties \& Stability of Junction Fixed Points}
Here we briefly review the properties and stabilities of the fixed points appearing in the main text. The discussion can be nicely captured by the technology known as ``delayed evaluation of boundary condition" (DEBC), which has been used for the spinless Y-junction problem.\cite{oshikawa2006junctions} Essentially, the DEBC lists possible local operators at the junction and investigates the scaling dimensions of the operators at a particular fixed point. The detailed procedure is not much important for us here. Note that even the exactly same UV operator can have the different scaling dimensions for the different IR fixed points. With the list of the operators and their scaling dimensions, we can now ask the stability of the fixed point. The fixed point is stable if the scaling dimensions of all the operators are bigger than 1. See the list of the operators for the spinless Y-junction in [Table \ref{Table_ScalingDim}]. Note again that the detailed forms of the operators and their names are not much important for our discussion below. The only important information for us is that we have a list of operators and their scaling dimensions. For those who want to learn more about DEBC, we refer the reader to the original Ref.[\onlinecite{oshikawa2006junctions}].

\begin{table}[ht]
\begin{tabular}{|c|c|c|c|c|}
\hline
                      & \textbf{$+$ cycle hopping}           & \textbf{$-$ cycle hopping}           & \textbf{Backscattering}          & \textbf{LL/RR hopping}                                              \\ \hline
\textbf{Operators}    & $\psi^{\dagger}_{R,j+1}\psi_{L,j}$ & $\psi^{\dagger}_{R,j}\psi_{L,j+1}$ & $\psi^{\dagger}_{R,j}\psi_{L,j}$ & $\psi^{\dagger}_{L,j+1}\psi_{L,j}$, or $\psi^{\dagger}_{R,j+1}\psi_{R,j}$ \\ \hline
\textbf{Disconnected} & $K$                                & $K$                                & 0                                & $K$                                                                 \\ \hline
\textbf{Connected}    & $1/3K$                             & $1/3K$                             & $1/K$                            & $4/3K$                                                              \\ \hline
\textbf{Chiral ($+$)}   & 0                                  & $4K/(1+3K^2)$                      & $4K/(1+3K^2)$                    & $4K/(1+3K^2)$                                                       \\ \hline
\textbf{Chiral ($-$)}   & $4K/(1+3K^2)$                      & 0                                  & $4K/(1+3K^2)$                    & $4K/(1+3K^2)$                                                       \\ \hline
\end{tabular}
\caption{The possible local operators at the Y-junction. $\psi^{\dagger}/\psi$ is the electronic creation/annihilation operator at the junction. Their scaling dimensions at each fixed points are listed. Here the two fixed points ``Chiral ($+$)" and ``Chiral ($-$)" are time-reversal partners.}
\label{Table_ScalingDim}
\end{table}

With the above in hand, we now analyze the stability and review some properties of the fixed points. 

\begin{itemize}
\item{\textbf{Disconnected Fixed Point:} Because the smallest scaling dimension of the operators at this fixed point is $K$, we find that the disconnected fixed point is stable if $1<K$. Exactly at this fixed point, the conductance vanishes, i.e. $G_{ij} =0$\cite{oshikawa2006junctions}, as the name of the fixed point suggests.}
\item{\textbf{Connected Fixed Point:} Because the smallest scaling dimension of the operators at this fixed point is $1/3K$, we find that the disconnected fixed point is stable if $1<1/3K$, or equivalently $0<K<1/3$. Precisely at this fixed point, the junction has the maximum conductance that the junction can have, i.e. $G_{ij}=(e^2/h)(2/3K)(3\delta_{ij}-1)$.\cite{oshikawa2006junctions}}
\item{\textbf{Chiral Fixed Point:} Because the smallest scaling dimension of the operators at this fixed point is $4K/(1+3K^2)$, we find that the disconnected fixed point is stable if $1<4K/(1+3K^2)$, or equivalently $1/3<K<1$. Exactly at this fixed point, the junction has the conductance $G_{ij}=(e^2/h) [(3\delta_{ij}-1)\pm \epsilon_{ij}/K ]\frac{2K}{1+3K^2}$.\cite{oshikawa2006junctions} Note that the chiral fixed point requires an explicit time-reversal symmetry breaking, e.g. magnetic field, to arise, although it could be infinitesimal.\cite{oshikawa2006junctions} In the main text, we consider the network subject under the uniform magnetic fields, when we discuss the NFL emerging from the chiral fixed point. }
\end{itemize}


\section{Conductance of Y-junction with Perturbations}
Here we present the calculation of the conductance of a single Y-junction at each fixed points appearing in our main text. We start with the free boundary action at the junction, which is just a sum of the contribution of each wire: 
\begin{equation}
S_0 = \frac{1}{\beta}\sum_{n} \sum_{j=1,2,3} \frac{|\omega_n|}{2K}|\theta_j (\omega_n)|^2
\text{ or }
S_0 = \frac{1}{\beta}\sum_{n} \sum_{j=1,2,3} \frac{K|\omega_n|}{2}|\phi_j (\omega_n)|^2, 
\end{equation}
with boundary conditions. We will use the rotated basis introduced in Ref.[\onlinecite{oshikawa2006junctions}]. Here we will present only the detailed calculations of the conductance at the disconnected fixed point. The calculation here can be straightforwardly generalizable to the other fixed points, too, and we sketch the calculations for the connected fixed point. 

\subsection{Disconnected Fixed Point} 
At the disconnected fixed point, we have the Neumann boundary condition for $\theta_j$ (superconducting phase) and the Dirichlet boundary condition for $\phi_j$ (charge density wave phase), i.e. $\partial_x \theta_j |_{x=0} = \text{Const.}_1,\: \phi_j (x=0) = \text{Const.}_2 $,
where $j=1,2,3$ and $\text{Const.}_{1,2}$ were set as zero for convenience. With these boundary conditions, we consider 
\begin{equation}
S_0 = \frac{1}{\beta}\sum^{\infty}_{n=-\infty} \sum_{j=1,2,3} \frac{|\omega_n|}{2K}|\theta_j (\omega_n)|^2,
\end{equation}
and hopping between the wires as a leading-order perturbation
\begin{equation}
H_B = \sum^{3}_{j=1} \Big[ (i\Gamma e^{i\tilde{\Phi} /3} \eta_{j+1} e^{i\sqrt{\pi}(\theta_j -\theta_{j-1})} +H.c.) -\frac{r}{2\sqrt{\pi}} \partial_x \theta_j (x)  \Big]\Big|_{x=0}, \label{C8}
\end{equation}
where $\tilde{\Phi}$ is the magnetic flux in a particular truncation of Y-junction in [\onlinecite{oshikawa2006junctions}], which we keep in our manuscript, too. $\eta_j$ is the Klein factor which has an anti-commuting statistics $\{\eta_i ,\eta_j \}=2\delta_{ij}$. Because of the boundary condition, the electronic fields are simplified to $\psi_R \sim e^{-i\sqrt{\pi}\theta}$ and $\psi_L \sim e^{-i\sqrt{\pi}\theta}$, i.e. the right mover and the left mover at the boundary are indistinguishable. In the rotated basis, we have
\begin{equation}
\begin{aligned}
S=S_0 +S_B = \frac{1}{\beta}\sum_{n} \sum^{2}_{j=1}  \frac{|\omega_n|}{2K} |\Theta_j (\omega_n)|^2+ i\Gamma e^{i\tilde{\Phi} /3}\int^{\beta}_{0} d\tau \sum^{3}_{j=1}  \eta_{j} e^{-i\sqrt{2\pi}\vec{K}_j \cdot \vec{\Theta} } +\text{H.c}. 
\end{aligned}
\end{equation}
One can expand the partition function in terms of $S_B$. The first-order term vanishes. In the second-order term, only $(i\rightarrow j)$ \& $(j\rightarrow i)$ pair terms can potentially be non-zero. Hence, we obtain 
\begin{equation}
\langle S^{2}_{B}  \rangle_0 = \Gamma^2 \int^{\beta}_{0} d\tau d\tau' \sum^{3}_{j=1} \langle T_{\tau} e^{-i\sqrt{2\pi}\vec{K}_j \cdot (\vec{\Theta}(\tau)-\vec{\Theta}(\tau'))} \rangle_0 + \text{H.c.}.
\end{equation}   
Note that the second-order term is $\tilde{\Phi}$ independent. However, the third-order term is $\tilde{\Phi}$-dependent: 
\begin{equation}
\begin{aligned}
\langle S^{3}_{B}  \rangle_0 =& 2\Gamma^3 e^{i\tilde{\Phi}} \int^{\beta}_{0} d\tau' d\tau'' d\tau''' \langle T_{\tau} e^{-i\sqrt{2\pi}(\vec{K}_1 \cdot \vec{\Theta}(\tau')+\vec{K}_2 \cdot \vec{\Theta}(\tau'')+\vec{K}_3 \cdot \vec{\Theta}(\tau'''))}\rangle_0 +\text{H.c.} 
+ \Big( \vec{K}_{j\rightarrow j+1} \text{Permutation} \Big). 
\end{aligned} 
\end{equation}
The below are the details of how we evaluated the correction to the conductance from the above integral formula. 

\subsubsection{Second-Order Term}
Let us first calculate the second-order perturbative correction to the conductance by following [\onlinecite{kane1992transport}].
\begin{equation}
\begin{aligned}
\langle S^{2}_{B} \rangle_0 &= 2\Gamma^2 \int^{\beta}_{0} d\tau d\tau' \sum^{3}_{j=1} P_{j}(\tau -\tau') \cos{(\sqrt{2}\vec{K}_j \cdot (\vec{a} (\tau)-\vec{a} (\tau')))},
\end{aligned}
\end{equation}
where $\vec{a}=(a_1,a_2)$ is the electromagnetic field coupling in the rotated basis of $A_{j=1,2,3}$. Here we have introduced the gauge fields to keep track of the electric currents. Note that the voltage $V_j = A_j t$. We calculate the current at each wire by differentiating the partition function with respect to the fields 
\begin{equation}
\begin{aligned}
I_j (\tau) &= \frac{\delta}{\delta A_j (\tau)} \Gamma^2 \int^{\beta}_{0} d\tau_1 d\tau_2 \sum^{3}_{k=1} P_{k}(\tau_1 -\tau_2) \cos{(\sqrt{2}\vec{K}_k \cdot (\vec{a} (\tau_1)-\vec{a} (\tau_2)))} .
\end{aligned}
\end{equation}
Note that the current $I_j$ is a function of the field $A_{j=1,2,3}$. We next perform the analytic continuation to calculate the correlation functions at the finite temperature.
\begin{equation}
I_j (t) = \Gamma^2 \Big( \frac{P_{j-2} (V_j -V_{j-1}) -P_{j-2} (-(V_j-V_{j-1}))}{i}-\frac{P_{j-1} (V_{j+1} -V_{j}) -P_{j-1} (-(V_{j+1}-V_{j}))}{i}\Big),
\end{equation}
where 
$
P_j (E) = -i\int^{\infty}_{0}dt (e^{iEt}-1) \Big(P^{>}_{j}(t)-P^{<}_{j}(t)\Big).
$
In the small voltage limit (linear response regime), $P_j (V_j) \simeq -P_j (-V_j)$, and $P_j (a+b) \simeq P_j (a)+P_j (b)$. Hence the current is
\begin{equation}
\begin{aligned}
I_j (t) &= \frac{4\pi e^2 \Gamma^2}{h} \Big( \frac{2P(V_j)}{iV_j}V_j -\frac{P(V_{j-1})}{iV_{j-1}}V_{j-1} -\frac{P(V_{j+1})}{iV_{j+1}}V_{j+1} \Big),
\end{aligned}
\end{equation}
where we put back the unit conductance $e^2 /\hbar$ into the expression. Performing the calculation for $P_j(V)$ explicitly,\cite{kane1992transport} we find 
\begin{equation}
(G_{\text{disconnected}})_{ij}=\Big( \frac{4\pi e^{2} \Gamma^2}{h} \tau^{2K }_{c} \frac{\pi^{2K-1}}{2} \frac{\Gamma(\frac{1}{2})\Gamma(K)}{\Gamma(\frac{1}{2}-K)} \Big) T^{2(K -1)} (3\delta_{ij}-1).
\end{equation}

\subsubsection{Third-Order Term}
We continue to the third-order correction for the completeness. The overall calculation is the same as the second order term above, and hence we only sketch the calculation. The third order term in the rotated basis is 
\begin{equation}
\begin{aligned}
& \langle S^{3}_{B} \rangle_0 = \Big[2\Gamma^3 e^{i\tilde{\Phi}} \int^{\beta}_{0} d\tau_1 d\tau_2 d\tau_3 \langle T_{\tau} e^{-i\sqrt{2\pi}(\vec{K}_1 \cdot \vec{\Theta}(\tau_1)+\vec{K}_2 \cdot \vec{\Theta}(\tau_2)+\vec{K}_3 \cdot \vec{\Theta}(\tau_3))} \rangle_0 \\
&\times e^{-i\sqrt{2}(\vec{K}_1 \cdot \vec{a}(\tau_1)+\vec{K}_2 \cdot \vec{a}(\tau_2)+\vec{K}_3 \cdot \vec{a}(\tau_3))} +\text{H.c.} \Big] + \Big( \vec{K}_1 \rightarrow \vec{K_2} \; \& \; \vec{K}_2 \rightarrow \vec{K_3} \; \& \; \vec{K}_3 \rightarrow \vec{K_1} \Big).   
\end{aligned} 
\end{equation}
As before, we track the current operator by performing the variation, i.e. 
$
I^{(3)}_{j} (\tau) = -\frac{1}{6} \frac{\delta \langle S^{3}_{B}\rangle_0 }{\delta A_j (\tau)}
$. 
In the small voltage limit, one can calculate the current along the $1$-wire, i.e. $I_{j=1} (\tau)$, for example. 
\begin{equation}
\begin{aligned}
I^{(3)}_{1} (\tau) &\simeq -12 \Gamma^3 \cos{\tilde{\Phi}}\int^{\beta}_{0} d\tau' d\tau'' P^{(3)}(\tau,\tau',\tau'')\Big( 2\sin{((A_1(\tau)-A_1(\tau''))}\\
&-\sin{((A_2(\tau)-A_2(\tau''))}-\sin{((A_3(\tau)-A_3(\tau''))} \Big) .
\end{aligned}
\end{equation}
Similarly, one can calculate $I^{(3)}_{2}(\tau)$ and $I^{(3)}_{3}(\tau)$. From this, we obtain the third order correction of the conductance tensor, which can be schematically written as 
$
(G^{(3)})_{ij} \propto \Gamma^3 \cos{\tilde{\Phi}} (3\delta_{ij}-1)
$.
To investigate the temperature dependence, we insert the correlation functions into the integral and extract the $T$-dependence as we did for the second-order term. We find that the correction has the following form  
\begin{equation}
\begin{aligned}
I^{(3)}_{1} (y) = -12 \Gamma^3 \cos{\tilde{\Phi}} \Big(\frac{\beta}{\pi}\Big)^2  \int^{\pi}_{0} dy_1 dy_2 \Big(\frac{1}{\beta}\Big)^{3K} \Big(\frac{(\pi \tau_c)^3 }{|y -y_1||y_1 -y_2||y_2 -y|} \Big)^{K} 
\Big[ \frac{2\beta}{i\pi}V_1 (y-y'')+(\cdots) \Big],
\end{aligned}
\end{equation}
where $y=\pi\tau/\beta$, $y_1=\pi\tau_1 /\beta$, and $y_2=\pi\tau_2 /\beta$. Collecting the terms linear in the applied voltage, we find the third order correction of the conductance tensor as 
$
(G^{(3)}_{\text{disconnected}})_{ij} \propto \Gamma^3 \cos{\tilde{\Phi}} \cdot T^{3(K -1)} \cdot (3\delta_{ij}-1), 
$
whose scaling in temperature is different from that of the second order term. 

\subsection{Connected Fixed Point} 
Our starting point is the conductance tensor exactly at the connected fixed point in the zero temperature at the limit of vanishing voltage.\cite{oshikawa2006junctions}
\begin{equation}
(G_{\text{Andreev}})_{ij}
=\frac{2}{3K}\frac{e^2}{h} (3\delta_{ij}-1).
\end{equation} 
This is the maximal conductance, which can be supported by a Y-junction \cite{oshikawa2006junctions}. Hence, at the finite temperature, whatever the perturbation we include, the conductance will decrease but in a power-law fashion. Hence, we expect to find a metallic behavior of the conductance in terms of the temperature.  

To explicitly calculate the temperature-depedent correction of the conductance, we again use the boundary action with the perturbation. The Dirichlet boundary condition is given for $\Theta_1, \Theta_2$ and the Neumann boundary condition for $\Theta_0$ at the connected fixed point. For this boundary condition, it is convenient to use  
\begin{equation}
S_0 = \frac{1}{\beta} \sum_{n} \sum_{j=1,2,3} \frac{K|\omega_n|}{2} |\phi_j (\omega_n)|^2 ,
\end{equation}
and +/- cycle hopping term as the perturbation [\onlinecite{oshikawa2006junctions}], which are the leading irrelevant term at the fixed point. Then, we can write down the total action coupled with the gauge field in the rotated basis as the following
\begin{equation}
\begin{aligned}
S&= \frac{1}{\beta}\sum_{n} \frac{K|\omega_n|}{2} \Big(|\Phi_1 (\omega_n)|^2 +|\Phi_2 (\omega_n)|^2 \Big)+\frac{1}{\beta}\sum_{n} \frac{|\omega_n|}{2K\pi} \Big(|a_1 (\omega_n)|^2 +|a_2 (\omega_n)|^2 \Big) \\
&+\Omega \sum^{3}_{j=1} \int^{\beta}_{0}d\tau \cos\Big(\sqrt{\frac{2}{3}}(\hat{z}\times \vec{K}_j )\cdot \Big(\sqrt{\pi}\vec{\Phi}(\tau)-\frac{1}{K}\vec{a}(\tau)  \Big) \Big). 
\end{aligned}
\end{equation}
Here the term $\propto \Omega$ is the perturbation. Note that the term describes so-called ``voltage-generating phase slips" (similar to the backscattering terms in the setup of \cite{kane1992transport}) and hence its effect is to decrease the overall conductivity \cite{Nayak1999}. We are particularly interested in the second-order perturbative correction: $-I^{(2)}(\tau) = (1/2)\delta \langle (S')^2 \rangle_0 / \delta a_j (\tau) $, where we have an explicit form 
\begin{equation}
\begin{aligned}
\langle (S')^2 \rangle_0 &= \frac{1}{2} \Omega^2  \int^{\beta}_{0}d\tau_1 d\tau_2 P(\tau_1 -\tau_2) \sum^{3}_{j=1} \cos \Big(\sqrt{\frac{2}{3}}\frac{1}{K} (\hat{z}\times\vec{K}_j )\cdot (\vec{a}(\tau_1)-\vec{a}(\tau_2))\Big),
\end{aligned}
\end{equation} 
with 
$
P(\tau_1 -\tau_2) = (\frac{\pi \tau_c /\beta}{\sin(\pi(\tau_1 -\tau_2 )/\beta)})^{\frac{2}{3 K}}
$. 
We calculate the current in the first wire, for example. In the small voltage limit, we find 
\begin{equation}
\begin{aligned}
-I^{(2)}_1 (\tau) &\simeq -\frac{1}{K^2 }\Omega^2 \int^{\beta}_{0} d\tau' P(\tau -\tau') \sin \Big(a_1(\tau)-a_1(\tau') \Big) .
\end{aligned}
\end{equation}
We next perform the analytic continuation and restore the unit conductance $e^2 /\hbar$ to find  
\begin{equation}
\begin{aligned}
G_{\mu=1,2} = \frac{2e^2}{Kh} - \Big( \frac{2\pi e^2 \Omega^2}{K^2 h}\tau^{2/3K}_{c}  \frac{\pi^{2/3K-1}}{2} \frac{\Gamma(\frac{1}{2})\Gamma(\frac{1}{3K})}{\Gamma(\frac{1}{2}-\frac{1}{3K})} \Big) T^{2/3K-2}.
\end{aligned}
\end{equation}
We perform the similar calculations for the other two wires to find 
\begin{equation}
(G_{\text{connected}})_{ij} =\Big(
\frac{2e^2}{3Kh} -  \frac{2\pi e^2 \Omega^2}{3K^2 h}\tau^{2/3K}_{c}  \frac{\pi^{2/3K-1}}{2} \frac{\Gamma(\frac{1}{2})\Gamma(\frac{1}{3K})}{\Gamma(\frac{1}{2}-\frac{1}{3K})}  T^{2/3K-2}
\Big) (3\delta_{ij}-1).
\end{equation}

\section{Structural Deformations of Network}
Here we will consider various structural deformation of the networks and their impacts on the 2D conductivity of the network. We will take the three representative deformations (which do not alter the connectivity of the wires), namely the uniform strain, missing Y-junctions, and the weak random strain. We will concentrate on the temperature scale where all the wires are safely in the 1D TLL regime and thus all the junctions in the network are safely described by the fixed points. Hence, the underlying junction conductance is permutation symmetric and isotropic.


\subsection{Uniformly Strained Networks}
\begin{figure}[ht]
\includegraphics[scale=0.5]{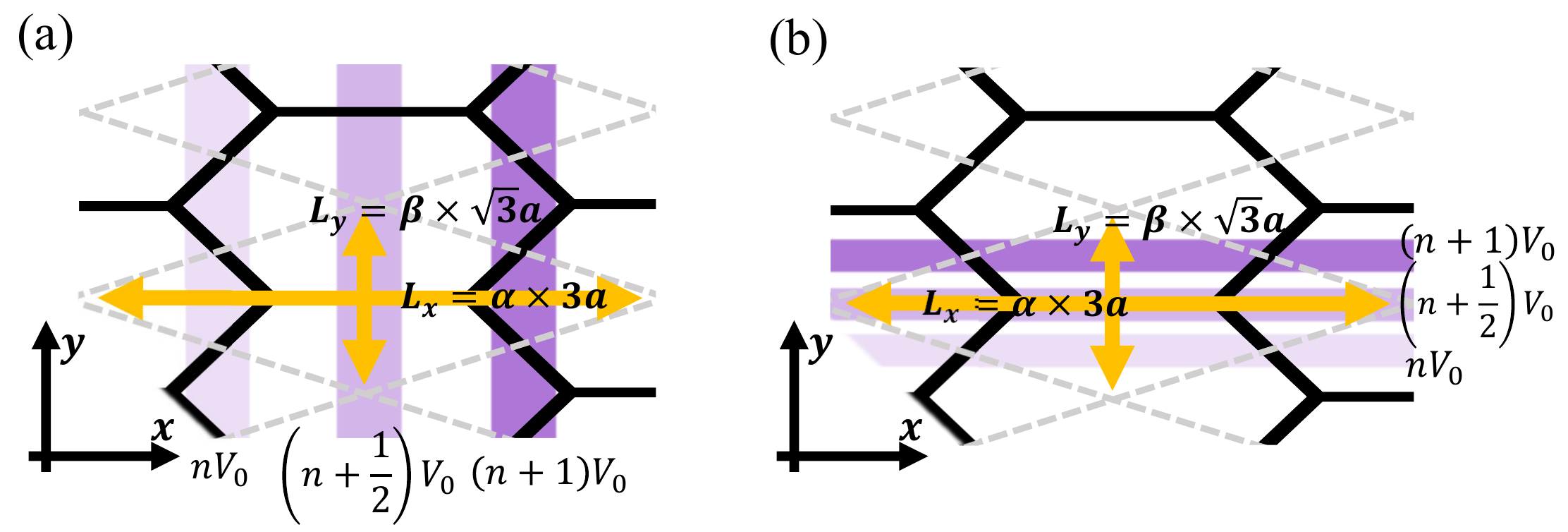}
\caption{Networks under the uniform strain. See the scaling factors compared to the perfect, isotropic network [Fig. \ref{General_Array}]. (a) Voltage drop is applied along the $x$-axis. (b) Voltage drop is applied along the $y$-direction. The yellow arrows indicate the size of the unitcell.}
\label{Fig_Strain}
\end{figure}

We consider the uniform strain on the network [Fig. \ref{Fig_Strain}]. Such deformation does not break the translation symmetry, which simplifies the problem. This case shows how the deformation can change 2D conductivity, despite the underlying junction conductance is permutation symmetric and isotropic.

Here the uniform strain introduced an anisotropic scaling of the unitcell for overall network, which will be parametrized by the two constants $(\alpha,\beta)$, i.e. $L_x \rightarrow \alpha L_x ,\; L_y \rightarrow \beta L_y$. We will focus on how these anisotropic factors appear in the 2D conductivity. 

For this, we first calculate the 2D conductivity when the electric field is along the y-direction: $\vec{E}= -2V_0 /\beta\sqrt{3}a \hat{y}$. See [Fig. \ref{Fig_Strain}(b)].  Given the voltage drops, the electric currents can be calculated, similar to the previous, isotropic uniform network case. $ I_1 = -(G_A +3G_S)V_0 /4,\; I_2 =(G_A -3G_S )V_0 /4,\; I_3 = -G_A V_0 /2$. From this, the current densities are obtained. 
\begin{equation}
\vec{J}= \Big(\frac{I_x}{\beta L_y },\frac{I_y}{\alpha L_x}\Big)= \Big(-\frac{G_A V_0}{\beta 2\sqrt{3}a},-\frac{G_S V_0}{\alpha 2 a}\Big).
\end{equation}
Note that the scaling factors appeared in the current densities and electric field. Finally, two components of the 2D conductivity are obtained by relating the current densities and electric fields to find $g_{yy}= (\beta/\alpha)\sqrt{3}G_S /4$ and $g_{xy} = G_A /4$. Similarly, when the electric field is applied along the x-axis [Fig. \ref{Fig_Strain}(a)], we obtain $g_{xx}= (\alpha/\beta) \sqrt{3}G_S /4$ and $g_{yx} = - G_A /4$. Hence, the full 2D conductivity of the network under the uniform strain is obtained as 
\begin{equation}
(g)_{ab} = 
\begin{pmatrix}
\frac{\alpha}{\beta} \frac{\sqrt{3}G_S}{4} & \frac{G_A}{4} \\
- \frac{G_A}{4} & \frac{\beta}{\alpha} \frac{\sqrt{3} G_S}{4}
\end{pmatrix}.
\end{equation}
Note that the rotational symmetry is explicitly broken if not $\alpha = \beta$. The key observation here is that the conductivity is almost identical to the original, isotropic network, except the geometric scaling factors $(\alpha,\beta)$. Hence, in this case, the power-law corrections (in temperature) to the 2D conductivities are identical to those of the perfect isotropic network. 

\subsection{Missing Y-Junctions}
\begin{figure}[ht]
\includegraphics[scale=0.5]{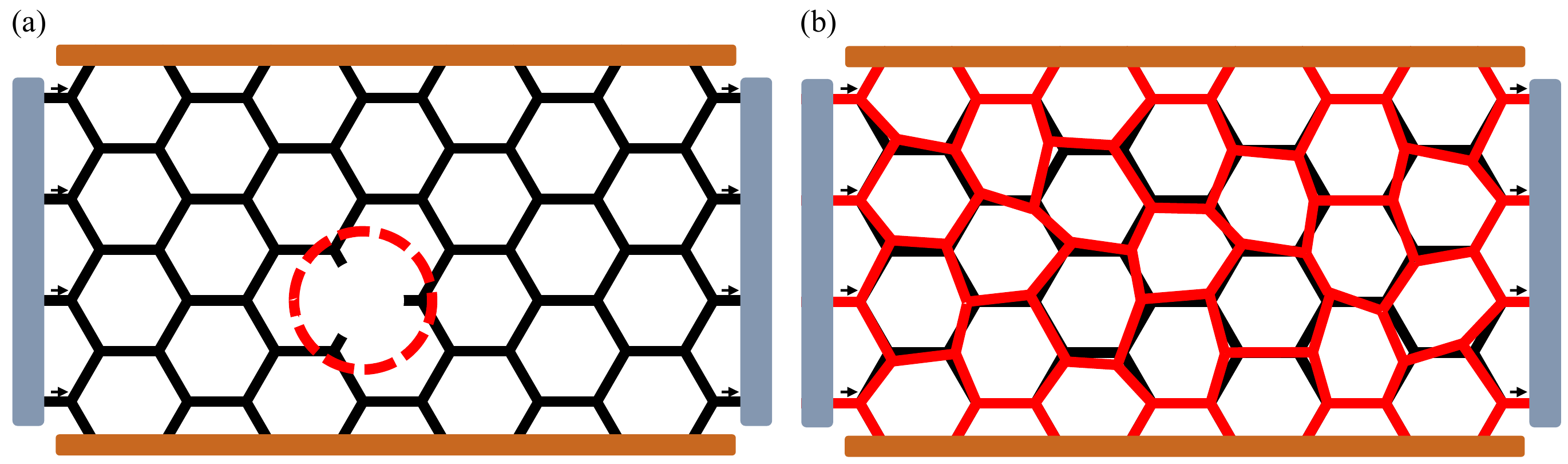}
\caption{(a) Network with a single Y-junction vacancy. (b) Randomly-strained network. The black bold lines represent the original, perfect network. The red lines represent the deformed network. Note that the wires at the boundaries are intact under the deformation and thus the deformed network has the same boundary conditions as the original perfect network. The gray bars at the left/right ends represent the source/sink of uniform charge currents. The brown bars at the top/bottom of the network represent the metals, imposing equipotential conditions on the wires in contact with them.}
\label{Fig_Impurity}
\end{figure}

Another interesting structural deformation of the network is the vacancy such as missing Y-junctions [Fig. \ref{Fig_Impurity}(a)]. Here we note that the missing Y-junctions are the junctions, where the currents cannot flow. Such junction is, by definition, identical to a junction at the disconnected fixed point although the fixed point is externally imposed by ``structure” instead of being emergent under the RG flow. This implies that the network with missing Y-junctions can be considered as a network where certain junctions are structurally imposed to be at the disconnected fixed point. From these, we can immediately deduce the impacts of the missing Y-junctions on the 2D conductivity. 

\begin{figure}[ht]
\includegraphics[scale=0.5]{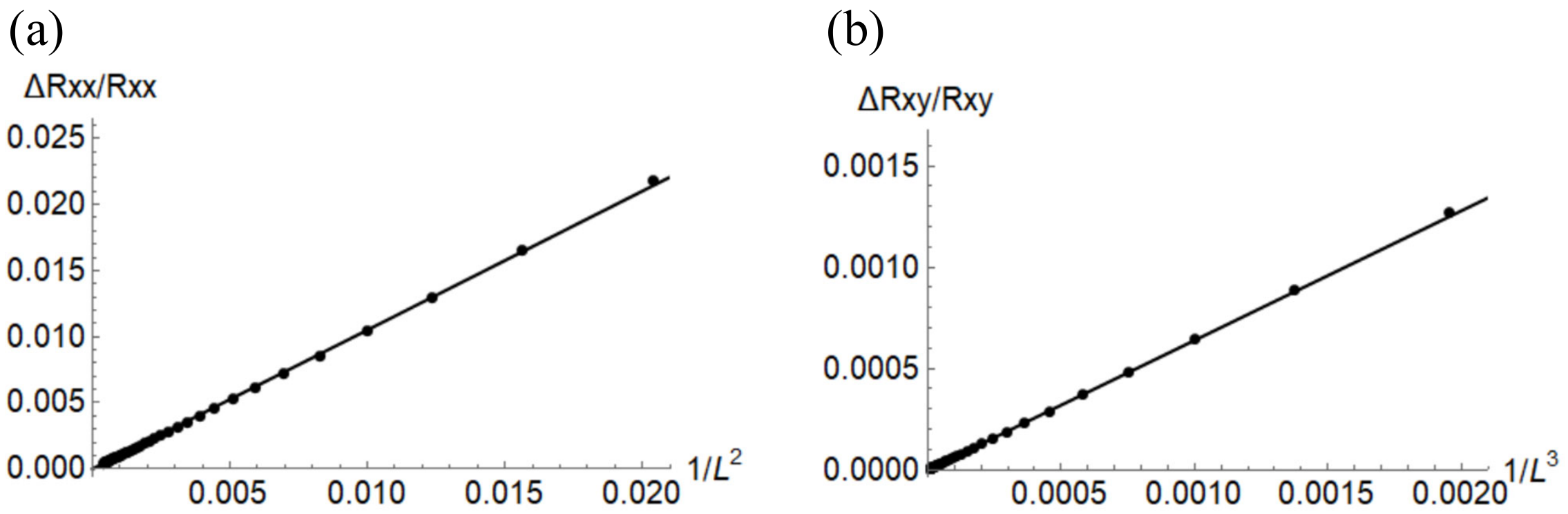}
\caption{The scaling of $|$ the correction to the resistivity due to a single missing Y-junction, i.e. $\Delta R_{ab}$ $|$/$|$(the resistivity of a perfect network $R_{ab}$) $|$ in terms of inverse of the system size $L$. (a) is the scaling of the correction for $R_{xx}$. (b) is the scaling of the correction to $R_{xy}$. It is clearly shown that, as the system size becomes larger, the resistivity approaches that of the clean, perfect network.}
\label{Fig_Vacancy_Data}
\end{figure}

\begin{itemize}
\item{If the ensemble of the configurations of missing Y-junctions do not break the crystalline symmetry, then the 2D conductivities (averaged over the ensemble) should remain isotropic, i.e. we expect $g_{xx} (T)=g_{yy} (T)$, $g_{xy} (T)= -g_{yx} (T)$.}
\item{Due to the appearance of the disconnected fixed point, the conductivity will receive an additional power-law correction in temperature, which is identical to that of the disconnected fixed point.}
\item{The correction to the conductivity of the perfect 2D network is expected to be small, if the density of the missing Y-junction is small. For instance, the corrections to the resistivity because of a single missing junction can be numerically simulated. See Fig.\ref{Fig_Vacancy_Data}. Here we took, as an example, that all the junctions (except the missing one) are at the chiral fixed points. There we explicitly see that the correction vanishes as the system size becomes large.}
\end{itemize}

\subsection{Randomly-Strained Networks}
Finally, we consider a network, which is randomly strained slightly [Fig. \ref{Fig_Impurity}(b)]. Here, the connectivity of wires and the overall honeycomb structure are intact although the shape of each honeycomb plaquette can be distorted arbitrarily (but weakly), i.e. all the junctions still connect the neighboring three wires. Because we focus on the temperature regime where all the wires are in the TLL regime, we expect that all the junctions are described by the same conductance tensor. With this, we find that the 2D conductivity of the deformed network will be the same as that of the perfect network after averaging over the configurations of such deformations. The proof consists of the two steps, which we explain below.

First, we consider a network with the weak random strain, which does not scale the overall size of the network. Such network can be generated by deforming the positions of junctions from the perfect network, while the position of the wires and junctions living at the boundary are fixed [Fig. \ref{Fig_Impurity}(b)]. Note that, these boundary wires are directly in contact with the external leads, which impose uniform voltages and insert uniform currents. This implies that the boundary conditions are intact despite of the deformation. Next we note that the equations that relate the current and voltages inside such deformed network remain the same as the perfect isotropic network. This is because of our assumption that we still consider the temperature regime where all the wires are governed by the identical TLL theory. This means that all the junctions in the deformed network are described by the same fixed-point conductance tensor, as in the perfect network. Hence, we find that the governing linear equations of currents and voltages are intact under the deformation and that the boundary conditions are also intact under the deformation. This immediately implies that solutions of the currents and voltages in this deformed network are identical to those of the perfect network. Hence, the conductivities of the deformed network are identical to those of the perfect network.

Next, we consider the deformations which accompany the global scaling of the networks, i.e. there are overall scaling factors in the system sizes $(L_x,L_y )\rightarrow(\alpha L_x,\beta L_y )$ with $(L_x, L_y)$ being the total length of the system. Then, the above considerations (uniformly strained case, and weak, random deformations without changing the global scales) immediately suggest that the conductance is given as below.
\begin{equation}
(g)_{ab} = 
\begin{pmatrix}
\frac{\alpha}{\beta} \frac{\sqrt{3}G_S}{4} & \frac{G_A}{4} \\
- \frac{G_A}{4} & \frac{\beta}{\alpha} \frac{\sqrt{3} G_S}{4}
\end{pmatrix}.
\end{equation}
The final step is then averaging over $(\alpha,\beta)$. For the reasonable random, weak distribution, e.g. box distribution of $\alpha \in (1-\epsilon_x ,1+\epsilon_x)$ and $\beta\in (1-\epsilon_y,1+\epsilon_y )$, we expect that the averaging over $(\alpha,\beta)$ will lead to the isotropic conductivities, which are the same as the perfect network.

\bibliography{Network}